\documentclass[12pt]{article}
\usepackage{graphicx}

\newcommand\pubnumber{SNSN-323-63}
\newcommand\pubdate{\today}

\usepackage{libertine}
\usepackage{TopAddendumModellingMain-defs}
\usepackage{multirow}
\usepackage{amsmath}
\usepackage{epstopdf}
\usepackage{cite}
\usepackage{graphicx}
\usepackage{caption}
\usepackage{subcaption}
\usepackage[left]{lineno}
\usepackage{multirow}
\usepackage{multirow,bigdelim}
\usepackage{blindtext}
\usepackage[T1]{fontenc}
\newlength{\bibitemsep}
\newlength{\bibparskip}
\let\oldthebibliography\thebibliography
\renewcommand\thebibliography[1]{%
  \oldthebibliography{#1}%
  \setlength{\parskip}{\bibitemsep}%
  \setlength{\itemsep}{\bibparskip}%
}

\def\institute{II. Physikalisches Institut, Georg-August-Universit\"at G\"ottingen, Germany}
\def\support{\footnote{E-mail: mamolla@cern.ch}}

\def\Title#1{\begin{center} {\Large #1 } \end{center}}
\def\Author#1{\begin{center}{ \sc #1} \end{center}}
\def\Address#1{\begin{center}{ \it #1} \end{center}}

\newcommand\pubblock{\rightline{\begin{tabular}{l} \pubnumber\\
         \pubdate  \end{tabular}}}
\newenvironment{Abstract}{\begin{quotation}  }{\end{quotation}}
\newenvironment{Presented}{\begin{quotation} \begin{center} 
             PRESENTED AT\end{center}\bigskip 
      \begin{center}\begin{large}}{\end{large}\end{center} \end{quotation}}

\def\beq{\begin{equation}}
\def\eeq#1{\label{#1}\end{equation}}
\def\eeqn{\end{equation}}

\def\beqa{\begin{eqnarray}}
\def\eeqa#1{\label{#1}\end{eqnarray}}
\def\eeqan{\end{eqnarray}}

\let\bar=\overbar

\def\Dslash{\not{\hbox{\kern-4pt $D$}}}
\def\dslash{\not{\hbox{\kern-2pt $\del$}}}

\def\msb{{\bar{\ssstyle M \kern -1pt S}}}

\begin{document}
\begin{titlepage}
\pubblock

\vfill
\Title{$t\bar{t}+V$ production in ATLAS}
\vfill
\Author{ Mar\'ia Moreno Ll\'acer\support, on behalf of the ATLAS Collaboration}
\Address{\institute}
\vfill
\begin{Abstract}
The latest ATLAS results for processes with a top quark pair and an associated vector boson are presented here.
The measurement of the production cross sections for these processes is important for the direct determination of the top quark couplings to gauge bosons and for constraints on new physics models, in particular for models which go beyond the Standard Model regarding the mechanism for the mass generation.
\end{Abstract}
\vfill
\begin{Presented}
$9^{th}$ International Workshop on Top Quark Physics\\
Olomouc, Czech Republic,  September 19--23, 2016
\end{Presented}
\vfill
\end{titlepage}
\def\thefootnote{\fnsymbol{footnote}}
\setcounter{footnote}{0}

\section{Introduction}
The productions of a boson ($X~=~\gamma$, $W$ or $Z$) in association with a top quark pair (referred to as $t\bar{t}\gamma$, $t\bar{t}W$ and $t\bar{t}Z$) are some of the main highlights of the physics programme of the LHC Run~2. The large center-of-mass energy available allows for the copious production of top quarks pairs in association with other particles at high transverse momentum enabling precise Standard Model (SM) measurements as well as probing many Beyond SM models that enhance rates for $t\bar{t}+X$ production.
 Measurements of the $t\bar{t}\gamma$ and $t\bar{t}Z$ processes allow for direct determination of the top quark couplings to bosons (top quark electric charge and weak isospin, respectively), while the $t\bar{t}W$ process is a SM source of same-sign dilepton events, which are the signature of many models of physics Beyond the SM. This report presents the latest experimental results from the ATLAS experiment~\cite{ATLAS} made since the Top2015 conference regarding these processes. In particular, the $t\bar{t}W$ and $t\bar{t}Z$ cross section measurements at $\sqrt{s}=$13~TeV~\cite{ATLAS_ttV13TeV} using the 2015 dataset are presented. Other results for these processes at the LHC at 8 or 13~TeV are documented in Refs.~\cite{ATLAS_ttV8TeV,CMS_ttV13TeV}.
\begin{figure}[h!]
  \centering
  \begin{subfigure}[a]{0.8\textwidth}
    \includegraphics[width=1.\textwidth]{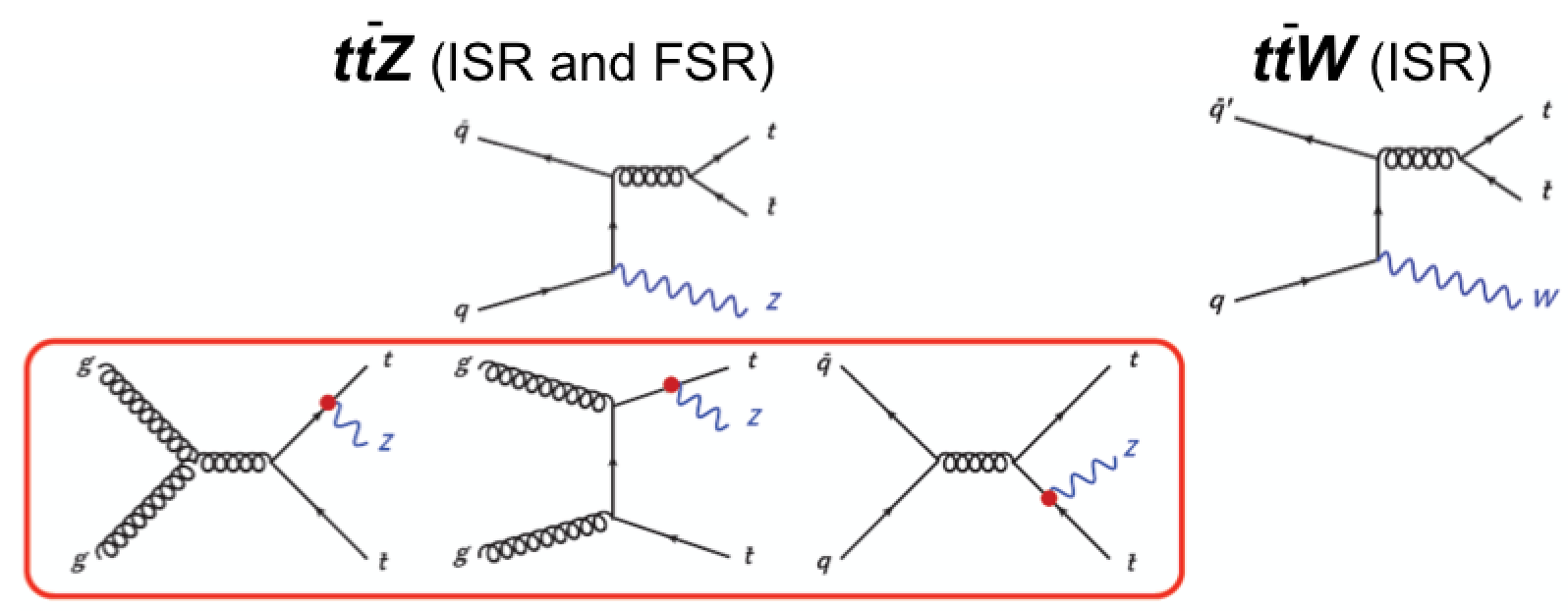}
  \end{subfigure}
  \begin{subfigure}[b]{1.\textwidth}
    \centering
    \includegraphics[width=0.45\textwidth]{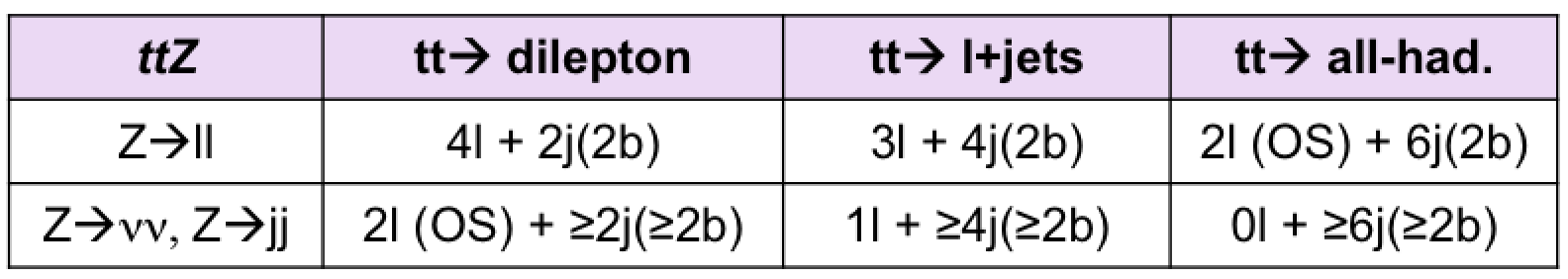}
    \includegraphics[width=0.45\textwidth]{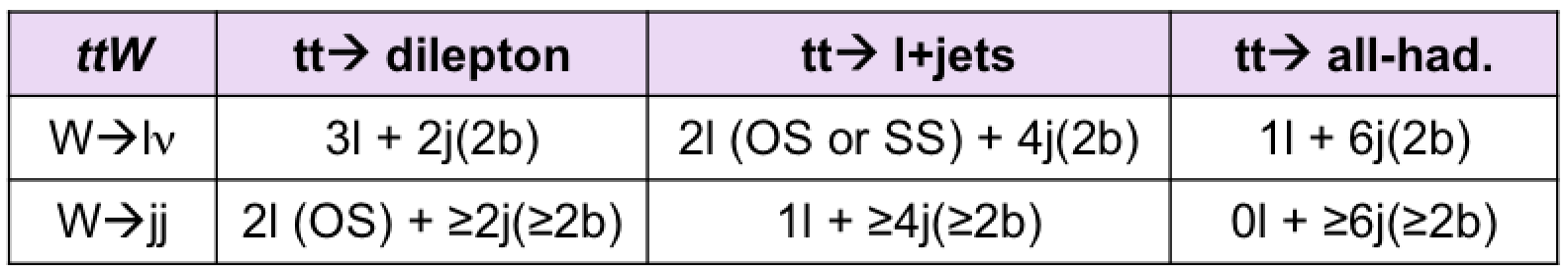}
  \end{subfigure}
  \caption{Example of leading-order Feynman diagrams for \ttZ~and \ttW~production.}
  \label{fig:ttVFeynmanDiagramsAndDecayModes}
\end{figure}
Fig.~\ref{fig:ttVFeynmanDiagramsAndDecayModes} shows some leading-order Feynman diagrams for \ttZ~and \ttW~production at the LHC. Depending on the decays of the top quarks, the $W$ and the $Z$ bosons, between zero and four prompt isolated leptons may be produced and thus different channels can be defined. So far, experimental analyses have focused on final states with two (same-sign or opposite-sign charge), three and four leptons.
The opposite-sign dilepton, trilepton and tetralepton channels are mostly sensitive to \ttZ~production, while the same-sign (SS) dilepton channel targets \ttW~production.
In order to enhance the sensitivity to the signal, each channel is further divided into multiple regions: ``signal regions'' (SR) where the \ttZ~/\ttW~processes give the dominant contribution and ``control regions'' (CR) where other better known SM processes are dominant. A simultaneous fit is performed to all signal and control regions in order to extract cross sections for \ttZ~and \ttW~production. Additionally, validation regions are defined to check that the background estimate agrees with the data, but are not used in the fit.

\section{\ttZ~and \ttW~cross section measurements at 13 TeV}

The latest cross section calculation at next-to-leading order (NLO) in QCD+EW for proton-proton collisions at a center-of-mass energy of $\sqrt{s}=13$~TeV predicts~\cite{HiggsYR4}:
\begin{eqnarray}
\sigma(t\bar{t}Z) & = & 839.3^{+9.6\%}_{-11.3\%}(\textrm{scale})\pm~2.8\%(\textrm{PDF})\pm~2.8\%(\alpha_S)~\textrm{fb}~~~\textrm{and} \nonumber \\
\sigma(t\bar{t}W) & = & 600.8^{+12.9\%}_{-11.5\%}(\textrm{scale})\pm~2.0\%(\textrm{PDF})\pm~2.7\%(\alpha_S)~\textrm{fb.} \nonumber 
\end{eqnarray}
\noindent
It has been calculated with \textsc{MadGraph5\_aMC@NLO} program~\cite{MG5_aMC@NLO,MC@NLO} using PDF4LHC15 PDF set and fixed QCD scales $\mu_0=\mu_R=\mu_F=m_\mathrm{top}~+~m_\mathrm{V}/2$ (with $V$~=$Z$, $W$).\\
For the first ATLAS 13~TeV results, a subset of the sensitive signal regions in the Run~1 analysis~\cite{ATLAS_ttV8TeV} is used (two same-sign muons, three and four leptons) and data collected in 2015 which corresponds to an integrated luminosity of 3.2~fb$^{-1}$ (with an uncertainty of $2.1\%$) are used. Only events collected using single-electron or single-muon triggers are accepted. Since they are almost fully efficient for leptons with $p_T >$25~GeV passing offline selections, this requirement is applied to (at least) the leading reconstructed lepton in all the channels.
Events must also have at least one reconstructed primary vertex. The three channels are further split and the full selection requirements for each of them are summarised in Table~\ref{tab:SRs}.

\begin{table}[h!]
\footnotesize
  \centering \renewcommand{\arraystretch}{1.2}
  \caption{\label{tab:SRs}Selection requirements in the signal regions defined in the two same-sign muons (top), trilepton (middle) and tetralepton (bottom) channels~\cite{ATLAS_ttV13TeV}. For the latter, all subleading leptons are required to satisfy $p_T>$7~GeV and the invariant mass of any OS lepton pair must be larger than 10~GeV.}
  \begin{tabular}{lccr@{}cc@{}lc}
    \hline
    Region & Leptons $p_T$ & $E_T^{\mathrm{miss}}$ & $H_{\mathrm{T,leptons+jets}}$~~&~~$n_{b{\mathrm{-tags}}}$\\
    \hline
    $2\mu$-SS & $\geq25$ GeV & $\geq40$ GeV & $\geq140$ GeV & 1\\
    \hline
  \end{tabular}

\begin{tabular}{l|c|c|c|c}
\hline
Variable  & 3$\ell$-Z-1b4j & 3$\ell$-Z-2b3j & 3$\ell$-Z-2b4 & 3$\ell$-noZ-2b\\
\hline
Leading lepton & \multicolumn{4}{c}{$p_T>25$ GeV} \\
Other leptons & \multicolumn{4}{c}{$p_T>20$ GeV} \\
Sum of lepton charges &  \multicolumn{4}{c}{$\pm1$} \\
$Z$-like OSSF pair & \multicolumn{3}{c|}{$|m_{\ell\ell} - m_Z| < 10$ GeV} & $|m_{\ell\ell} - m_Z| > 10$ GeV\\ 
$n_{\mathrm{jets}}$ & $\ge 4$ & $3$ &  $\ge 4$ & $\ge2$ and $\le4$\\
$n_{b{\mathrm{-tags}}}$ & $1$ & $\ge2$ & $\ge2$ & $\ge2$\\
\hline
\end{tabular}

  \begin{tabular}{lccr@{}cc@{}lc}
    \hline
    Region & Second lepton pair ($Z_2$) &  $p_{T,3}+p_{T,4}$ && $|m_{Z_{2}} - m_Z| $ & $E_T^{\mathrm{miss}}$ && $n_{b{\mathrm{-tags}}}$\\
    \hline
    4$\ell$-DF-1b & $e^{\pm}\mu^{\mp}$ & $>35$ GeV &&-&-&& 1\\
    4$\ell$-DF-2b & $e^{\pm}\mu^{\mp}$ & -&&-&-&& $\ge2$\\
    4$\ell$-SF-1b & $e^{\pm}e^{\mp},\mu^{\pm}\mu^{\mp}$ & $>25$ GeV & \begin{tabular}{@{}r@{}} \ldelim\{{2}{2ex} \\ \\ \end{tabular} &
    \begin{tabular}{c}$>10$ GeV\\ $<10$ GeV\end{tabular} & \begin{tabular}{c}$>40$ GeV\\ $>80$ GeV\end{tabular} &\begin{tabular}{@{}l@{}} \rdelim\}{2}{2ex} \\ \\ \end{tabular} & 1\\
    4$\ell$-SF-2b & $e^{\pm}e^{\mp},\mu^{\pm}\mu^{\mp}$ &-&\begin{tabular}{@{}r@{}} \ldelim\{{2}{2ex} \\ \\ \end{tabular} &
    \begin{tabular}{c}$>10$ GeV\\ $<10$ GeV\end{tabular} & \begin{tabular}{c}-\\ $>40$ GeV\end{tabular} &\begin{tabular}{@{}l@{}} \rdelim\}{2}{2ex} \\ \\ \end{tabular} & $\ge2$ \\
    \hline
  \end{tabular}
\end{table}

\begin{figure}[h!]
  \centering
  \begin{subfigure}[b]{1.\textwidth}
    \includegraphics[width=0.32\textwidth]{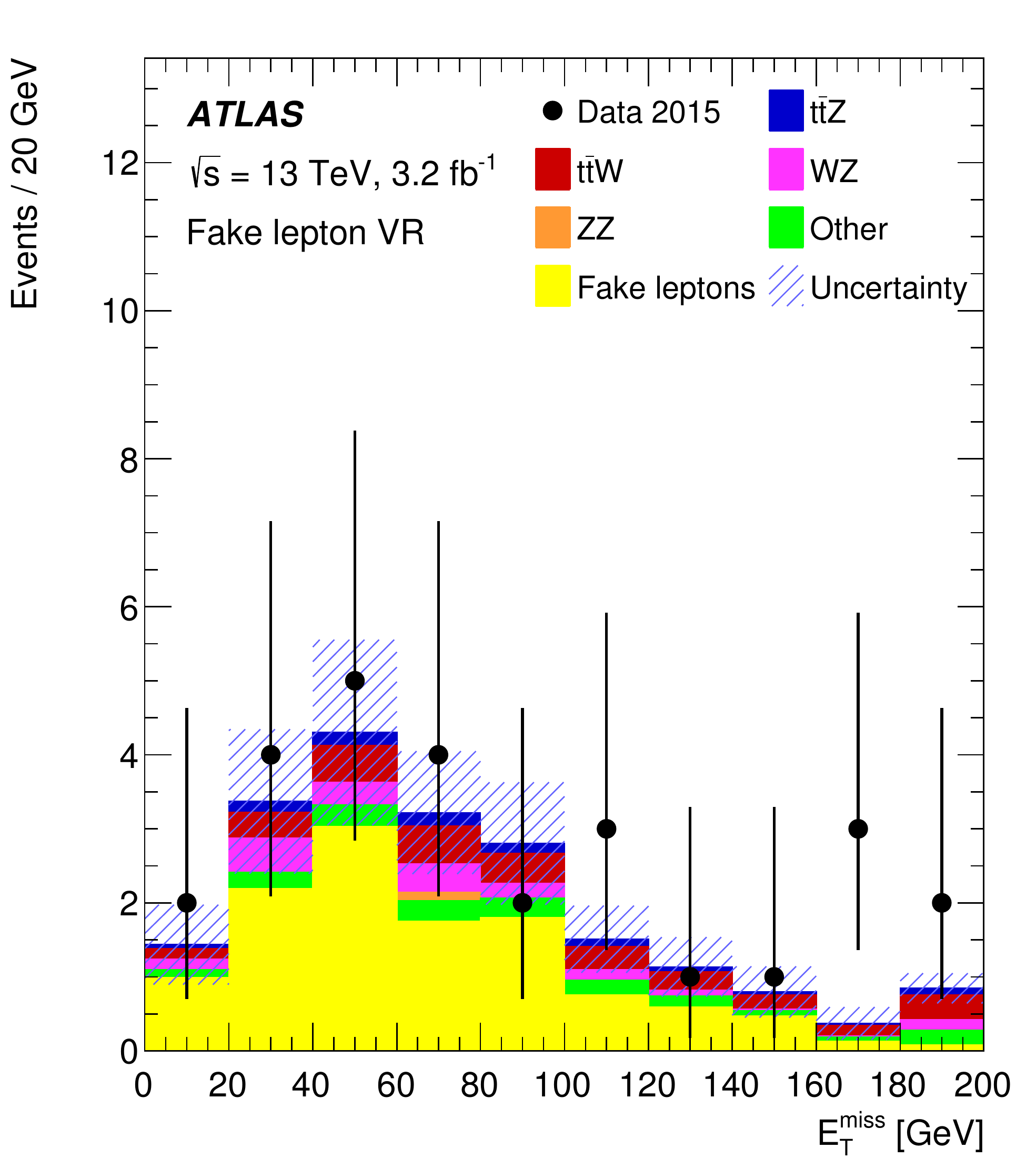}
    \includegraphics[width=0.32\textwidth]{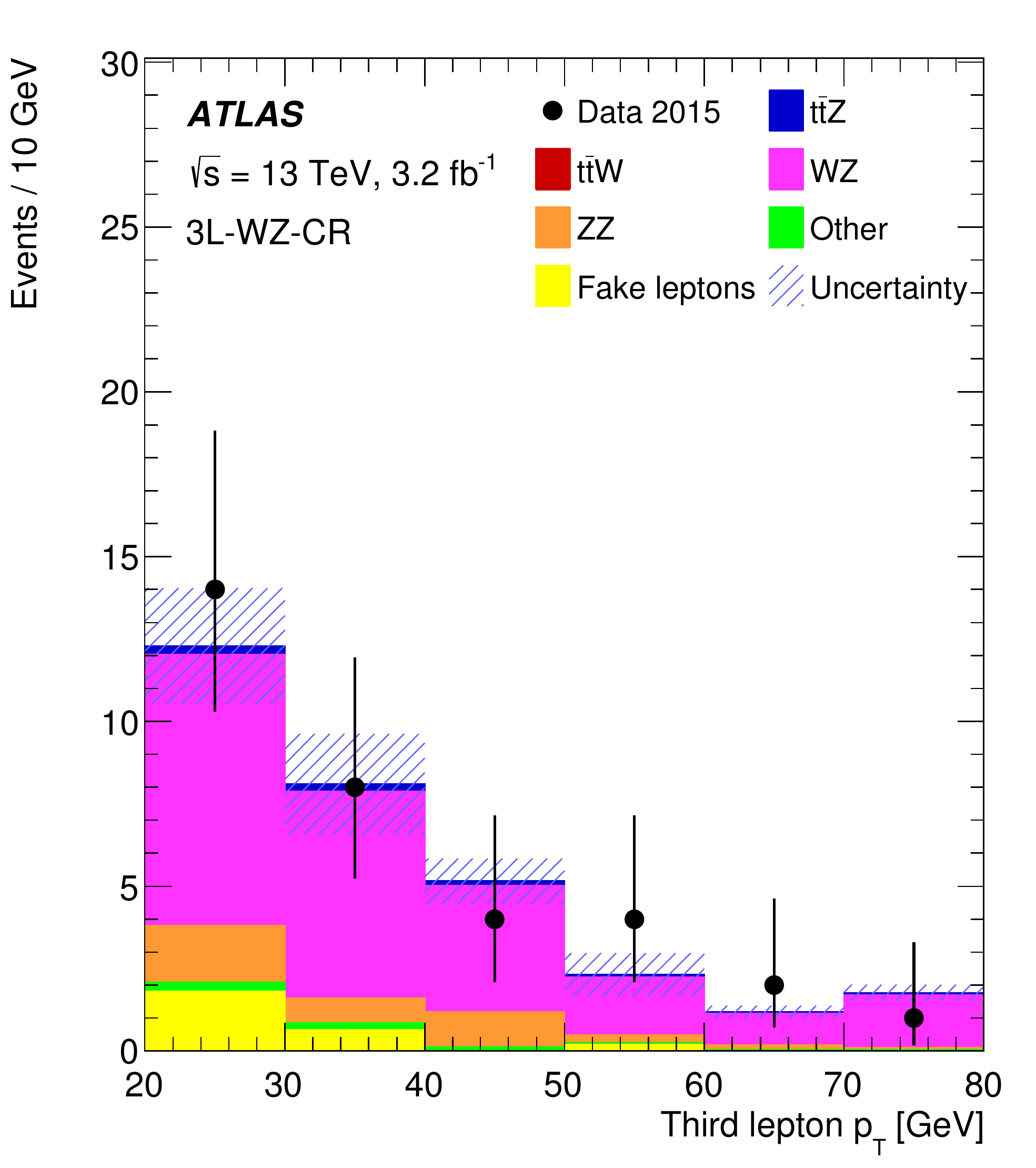}
    \includegraphics[width=0.32\textwidth]{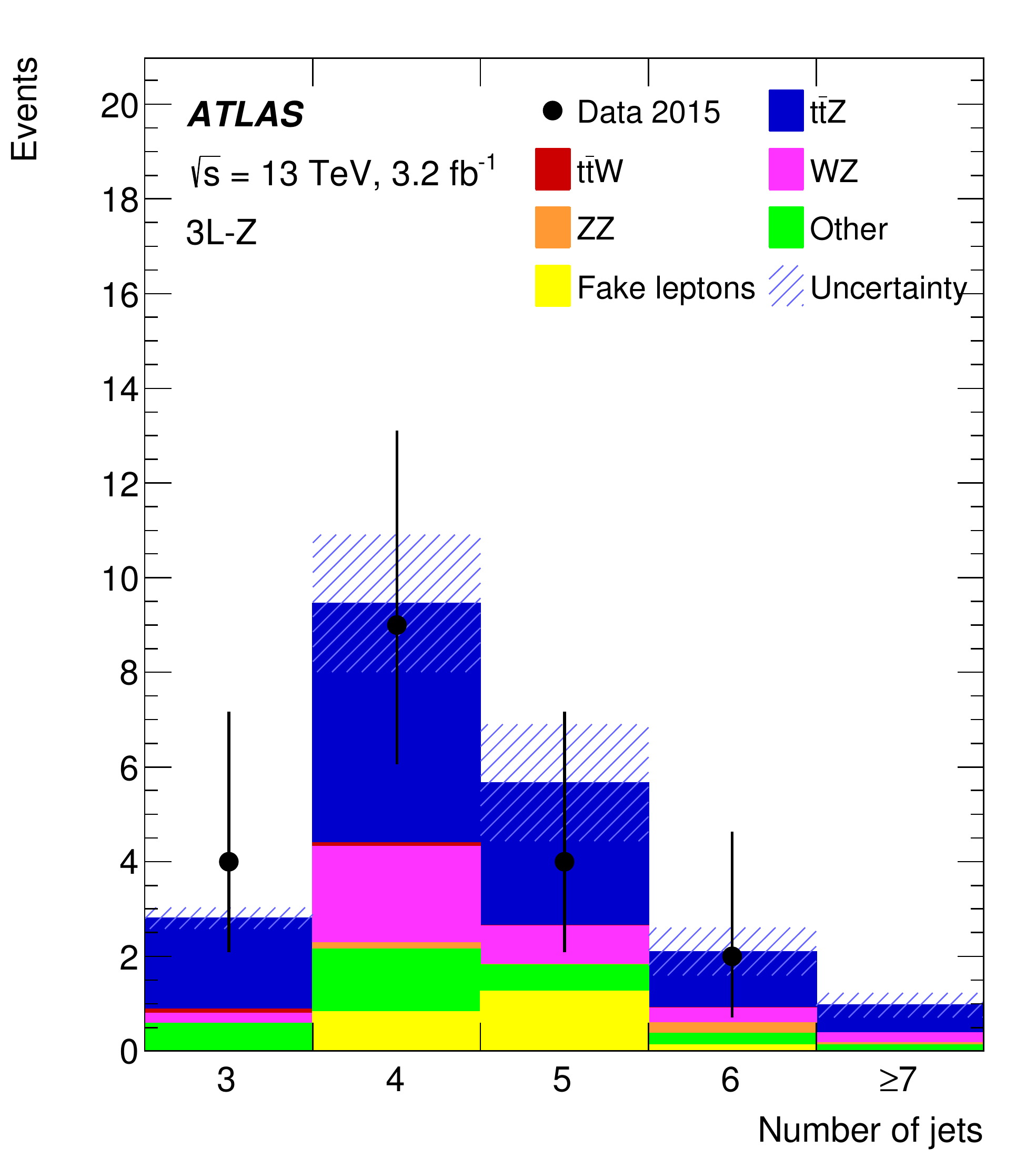}
  \end{subfigure} 
\caption{Distribution of the missing transverse momentum in the $2\mu$-SS validation region (left), transverse momentum of the third-lepton $p_T$ in the 3$\ell$-WZ-CR control region (middle) and number of jets (right) for events in the trilepton signal regions~\cite{ATLAS_ttV13TeV}. The distributions are shown before the fit. The shaded bands include the total uncertainties.}
  \label{fig:CRandSRplots}
\end{figure}

Monte Carlo simulation samples are used to model the expected signal and background distributions in the different analysis regions. 
Background events containing prompt leptons are modelled by simulation. The normalisations for the $WZ$ and $ZZ$ processes, main backgrounds in the channels with three and four prompt leptons, are taken from data control regions and 
included in the fit. The yields in these control regions are extrapolated to the signal regions. 
In the control region defined in the three lepton channel (3$\ell$-WZ-CR), exactly three leptons are required, at least one pair of which must be an opposite-sign same-flavour pair with an invariant mass within 10~GeV of the $Z$ boson mass and there must be exactly three jets, none of which pass the $b$-tagging requirement. Similarly, the four lepton channel control region (4$\ell$-ZZ-CR) is defined to have exactly four reconstructed leptons, two opposite-sign same-flavour pairs with an invariant mass within 10~GeV of the $Z$ boson mass and $E_T^{\mathrm{miss}}<$40~GeV. 
Background sources involving one or more fake leptons are modelled using data events from control regions.
For the same-sign dimuon and the trilepton analyses, the fake lepton background is estimated using the matrix method~\cite{ATLAS_MM}.
The control regions are defined in dilepton events, separately for $b$-tagged and $b$-vetoed events to take into account whether the source of the fake lepton is a light- or a heavy-flavour jet. The lepton efficiencies are extracted in a likelihood fit assuming Poisson statistics and that events with two fake leptons are negligible. The real lepton efficiencies are measured in inclusive opposite-sign events, and fake lepton efficiencies in events with same-sign leptons and $E_T^{\mathrm{miss}}>$40~GeV (for $b$-tagged events $E_T^{\mathrm{miss}}>$20~GeV), after subtracting the estimated contribution from events with misidentification of the charge of a lepton, and excluding the same-sign dimuon signal region. For the four leptons channel, the contribution from backgrounds containing fake leptons is estimated from simulation and corrected with scale factors determined in control regions.
The dominant background in the $2\mu$-SS and 3$\ell$-noZ-2b regions arises from events containing fake leptons. In order to check that this background is properly estimated, validation regions are defined based on the signal regions selection but relaxing some of the cuts. 
Fig.~\ref{fig:CRandSRplots} shows some distributions in the $2\mu$-SS validation region, and the $WZ$ control region and signal regions defined in the three channels. The expected event yields for signal and backgrounds, and the observed data in all control and signal regions used in the fit are shown in Table~\ref{tab:yields}.\\

\begin{table}[htbp]
\scriptsize
  \centering \renewcommand{\arraystretch}{1.2}
  \caption{\label{tab:yields} Expected and observed event yields~\cite{ATLAS_ttV13TeV}. The quoted uncertainties in the former represent systematic uncertainties. The $tZ$, $WtZ$, \ttH, three and four top quark processes are
    denoted as $t+X$. The processes $WZ$, $ZZ$, $H \to ZZ$ (ggF and VBF), $HW$ and $HZ$, as well as electroweak processes involving the vector boson scattering diagram, are denoted as `Bosons'.}
  \begin{tabular}{%
c|
r@{\,}@{$\pm$}@{\,}l
r@{\,}@{$\pm$}@{\,}l
r@{\,}@{$\pm$}@{\,}l
r@{\,}@{$\pm$}@{\,}l
|r@{\,}@{$\pm$}@{\,}l
r@{\,}@{$\pm$}@{\,}l
|r
}
    \hline
    Region & \multicolumn{2}{c}{$t+X$} & \multicolumn{2}{c}{Bosons} & \multicolumn{2}{c}{Fake leptons} & \multicolumn{2}{c|}{Total bkg.} & \multicolumn{2}{c}{\ttW} & \multicolumn{2}{c|}{\ttZ} & Data \\
    \hline
    3$\ell$-WZ-CR&   $0.52$ & $0.13$&    $26.9$ & $2.2$&   $2.2$ & $1.8$&    $29.5$ & $2.8$& $0.015$ & $0.004$&   $0.80$ & $0.13$&    33\\
    4$\ell$-ZZ-CR&   \multicolumn{2}{c}{$<0.001$} &    $39.5$ & $2.6$&   $1.8$ & $0.6$& $41.2$ & $2.7$&    \multicolumn{2}{c}{$<0.001$} &    $0.026$ & $0.007$&    39\\
    \hline
    $2\mu$-SS&    $0.94$ & $0.08$&    $0.12$ & $0.05$&    $1.5$ & $1.3$&    $2.5$ & $1.3$&   $2.32$ & $0.33$&    $0.70$ & $0.10$&    9\\
    3$\ell$-Z-2b4j&    $1.08$ & $0.25$&    $0.5$ & $0.4$&    \multicolumn{2}{c}{$<0.001$} & $1.6$ & $0.5$&   $0.065$ & $0.013$&    $5.5$ & $0.7$&    8\\
    3$\ell$-Z-1b4j&    $1.14$ & $0.24$&    $3.3$ & $2.2$&    $2.2$ & $1.7$&    $6.7$ & $2.8$&   $0.036$ & $0.011$&    $4.3$ & $0.6$&    7\\
    3$\ell$-Z-2b3j&    $0.58$ & $0.19$&    $0.22$ & $0.18$&    \multicolumn{2}{c}{$<0.001$} &    $0.80$ & $0.26$&    $0.083$ & $0.014$&    $1.93$ & $0.28$&    4\\
    3$\ell$-noZ-2b&    $0.95$ & $0.11$&    $0.14$ & $0.12$&    $3.6$ & $2.2$&    $4.7$ & $2.2$&   $1.59$ & $0.28$&    $1.45$ & $0.20$&    10\\
    4$\ell$-SF-1b&    $0.212$ & $0.032$&    $0.09$ & $0.07$&    $0.113$ & $0.022$& $0.42$ & $0.08$&   \multicolumn{2}{c}{$<0.001$} &    $0.66$ & $0.09$&    1\\
    4$\ell$-SF-2b&    $0.121$ & $0.021$&    $0.07$ & $0.06$&    $0.062$ & $0.012$& $0.25$ & $0.07$&   \multicolumn{2}{c}{$<0.001$} &    $0.63$ & $0.09$&    1\\
    4$\ell$-DF-1b&    $0.25$ & $0.04$&    $0.0131$ & $0.0032$&    $0.114$ & $0.019$& $0.37$ & $0.04$&   \multicolumn{2}{c}{$<0.001$} &    $0.75$ & $0.10$&    2\\
    4$\ell$-DF-2b&    $0.16$ & $0.05$&    \multicolumn{2}{c}{$<0.001$} &    $0.063$ & $0.013$&   $0.23$ & $0.05$&    \multicolumn{2}{c}{$<0.001$} &    $0.64$ & $0.09$&    1\\ 
    \hline
  \end{tabular}
\end{table}

\begin{figure}
  \begin{minipage}[c]{0.5\textwidth}%
    \centering
    \includegraphics[width=1.\textwidth]{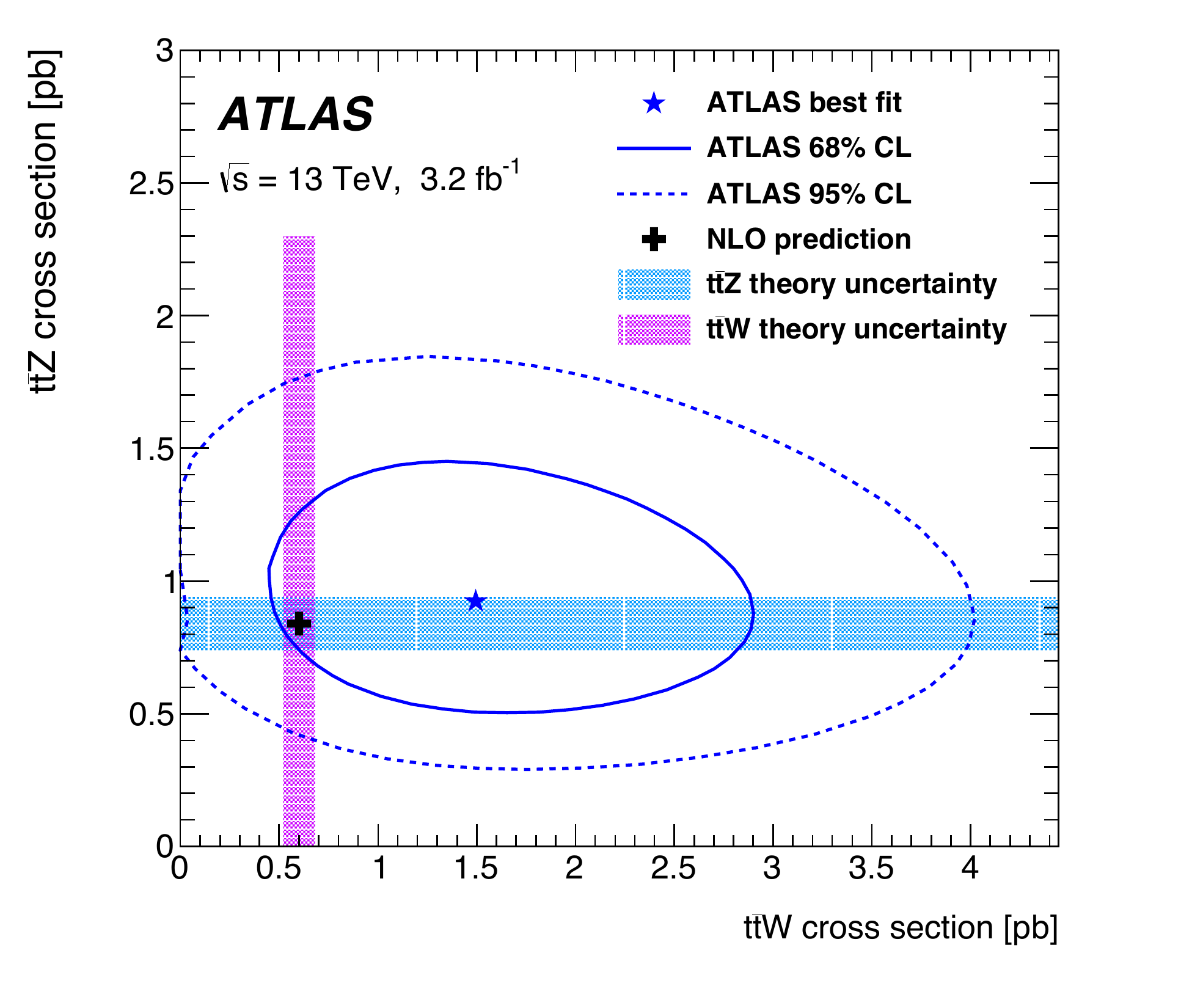}
  \end{minipage}
  \qquad
  \parbox{0.5\textwidth}{
    \begin{scriptsize}
      \begin{tabular}{lcc}
        \hline
        Uncertainty                 &   $\sigma_{\ttZ}$ & $\sigma_{\ttW}$ \\
        \hline
        Luminosity                    & 2.6\% & 3.1\%  \\
        Reconstructed objects         & 8.3\% & 9.3\%  \\
        Backgrounds from simulation   & 5.3\% & 3.1\%  \\
        Fake leptons and charge misID & 3.0\% & 18.7\%  \\
        Signal modelling   & 2.3\% & 4.2\%  \\
        \hline
        Total systematic            & 11\% & 22\% \\
        Statistical                 & 31\% & 48\% \\
        \hline
        Total                       & 32\% & 53\% \\
        \hline
      \end{tabular}
      \label{tab:syst}
    \end{scriptsize}
  }  \label{tab:syst}
    \caption{Result of the simultaneous fit to the \ttZ~and \ttW~cross sections along with the 68\% and 95\% confidence level (CL) contours (left) and breakdown of uncertainties (right)~\cite{ATLAS_ttV13TeV}. The shaded areas in the left plot correspond to the theoretical uncertainties in the SM predictions, and include QCD scales and PDF uncertainties.\label{fig:2Dplot_tab:syst}}
\end{figure}

The \ttZ~and \ttW~cross sections are determined using a binned maximum profile likelihood fit, where systematic uncertainties are allowed to vary as nuisance parameters. None of the uncertainties are
found to be significantly constrained or pulled from their initial values. The results of the fit are $\sigma_{\ttZ} = 0.92 \pm 0.29~\text{(stat)} \pm 0.10~\text{(syst)}$~pb and $\sigma_{\ttW} = 1.50 \pm 0.72~\text{(stat)} \pm 0.33~\text{(syst)}$~pb with a
correlation coefficient of $-0.13$ and are shown on the left side of Fig.~\ref{fig:2Dplot_tab:syst}. The leading and total uncertainties are listed on the right side.
For both processes, the precision of the measurement is dominated by statistical uncertainties.
The observed (expected) significances are $3.9\sigma$ ($3.4\sigma$) and $2.2\sigma$ ($1.0\sigma$) over the background-only
hypothesis for the \ttZ~and \ttW~processes, respectively.

\section{Conclusions}
The increase of energy at the LHC Run~2 allows to measure $t\bar{t}+X$ processes with large statistics and reaching kinematic regions never tested before. Here, the first measurement of \ttZ~and \ttW~production cross sections at 13~TeV performed by the ATLAS Collaboration using 3.2~fb$^{-1}$ of data is presented. The result has a relative uncertainty of $\mathcal{O}(30\%-50\%)$ and is limited by the statistical uncertainty, which can be vastly improved with the higher integrated luminosity now available (35~fb$^{-1}$).



\begin{thebibliography}{99}
\bibitem{ATLAS} ATLAS Collaboration, \emph{JINST} \textbf{3} S08003 (2008).
\bibitem{ATLAS_ttV13TeV} ATLAS Collaboration, arXiv:hep-ph/1609.01599
\bibitem{ATLAS_ttV8TeV} ATLAS Collaboration, \emph{JHEP} \textbf{11}, 172 (2015).
\bibitem{CMS_ttV8TeV} CMS Collaboration, \emph{Eur. Phys. J.} \textbf{C74}, 3060 (2014).
\bibitem{CMS_ttV13TeV} CMS Collaboration, CMS-PAS-TOP-16-017, 2016
\bibitem{HiggsYR4} D. de Florian \emph{et al.}, arXiv:hep-ph/1610.07922
\bibitem{MG5_aMC@NLO} S. Frixione, \emph{JHEP} \textbf{06}, 184 (2015).
\bibitem{MC@NLO} J. Alwall \emph{et al.}, \emph{JHEP} \textbf{07}, 079 (2014).
\bibitem{ATLAS_MM} ATLAS Collaboration, \emph{Eur. Phys. J.} \textbf{C71}, 1577 (2011).
\end{thebibliography}
\end{document}